\documentclass[manuscript,screen,nonacm,table]{acmart}

\setcopyright{acmcopyright}
\copyrightyear{2018}
\acmYear{2018}
\acmDOI{10.1145/1122445.1122456}

\acmConference[Woodstock '18]{Woodstock '18: ACM Symposium on Neural
  Gaze Detection}{June 03--05, 2018}{Woodstock, NY}
\acmBooktitle{Woodstock '18: ACM Symposium on Neural Gaze Detection,
  June 03--05, 2018, Woodstock, NY}
\acmPrice{15.00}
\acmISBN{978-1-4503-XXXX-X/18/06}

\usepackage{enumitem}
\usepackage{listings}
\usepackage{tabularx, makecell}
\usepackage{subcaption}
\usepackage[percent]{overpic}

\definecolor{gray}{rgb}{0.4,0.4,0.4}
\definecolor{darkblue}{rgb}{0.0,0.0,0.6}
\definecolor{cyan}{rgb}{0.0,0.6,0.6}

\definecolor{light-gray}{gray}{0.75}
\lstdefinelanguage{XML}
{
  morestring=[b]",
  morestring=[s]{>}{<},
  morecomment=[s]{<?}{?>},
  stringstyle=\color{black},
  identifierstyle=\color{darkblue},
  keywordstyle=\color{cyan},
  morekeywords={xmlns,version,type}
}

\newcommand{\SysNoSpace}{LaserSVG}
\newcommand{\Sys}{\SysNoSpace\ }

\newcommand{\thing}[1]{\href{https://www.thingiverse.com/thing:#1}{#1}}

\def\plaintitle{LaserSVG: Responsive Laser-Cutter Templates}

\def\plainkeywords{Laser Cutter; File Format; Vector; Fabrication; Parametric Design, SVG.}

\begin{document}

\title{\plaintitle}

\author{Florian Heller}
\orcid{0000-0002-3265-3570}
\affiliation{%
  \institution{Hasselt University — tUL — Flanders Make}
  \city{Diepenbeek}
  \country{Belgium}}
\email{florian.heller@gmail.com}
\author{Raf Ramakers}
\orcid{0000-0001-6466-0663}
\affiliation{%
  \institution{Hasselt University — tUL — Flanders Make}
  \city{Diepenbeek}
  \country{Belgium}}
\email{raf.ramakers@uhasselt.be}
\author{Kris Luyten}
\orcid{0000-0002-4194-1101}
\affiliation{%
  \institution{Hasselt University — tUL — Flanders Make}
  \city{Diepenbeek}
  \country{Belgium}}
\email{kris.luyten@uhasselt.be}

\renewcommand{\shortauthors}{Heller et al.}

\begin{teaserfigure}
	\centering
	\includegraphics[width=\textwidth]{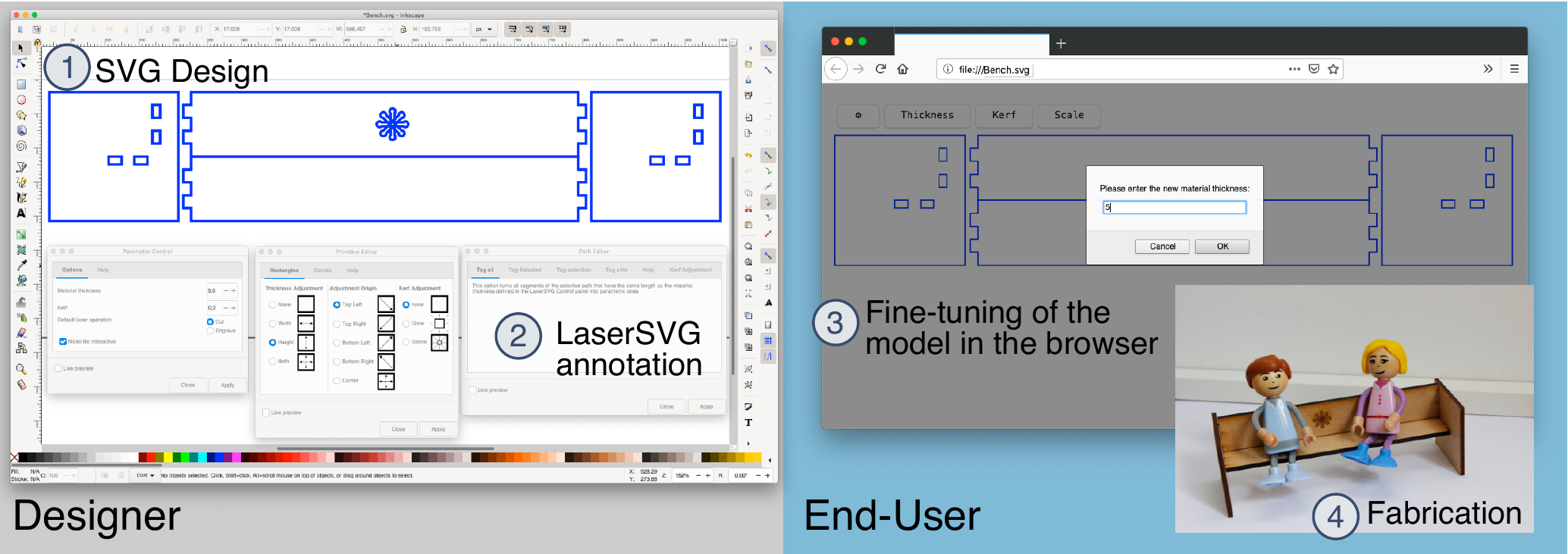}
	\caption{The designer draws or loads the original model of a bench as a regular SVG-file in Inkscape where she turns it into a \Sys template by tagging the edges that represent material thickness. When an end-user loads this template in a web-browser, a mini-menu generated by our JavaScript  library embedded into the template allows her to adjust the design to a specific material thickness or to scale it without affecting the fit of the elements.}~\label{fig:walkthrough}
	\vspace{-1.5em}
\end{teaserfigure}

\begin{abstract}
Laser cutters take vector data for the shapes they cut or engrave as input, however, re-using a given design with different material or on a different machine requires adaptation of the template. 
Unfortunately, vector drawings lack the semantic information required for an automated adjustment to new parameters, making the manual adjustment a tedious and error-prone process for end-users.
We present \SysNoSpace, a standard-compliant vector-based file format, software library, and authoring tool to specify, generate, exchange and re-use responsive laser-cutting templates.
With \SysNoSpace, designers can easily turn their vector-drawings into parametric templates that end-users can easily adjust to new materials or production parameters. 
Our tools provide various functions for parametric design that allows end-users and designers to adapt objects while ensuring overall consistency of the results. 
\end{abstract}

%
%

\ccsdesc[300]{Computing methodologies~Graphics file formats}
\ccsdesc[100]{Software and its engineering~Markup languages}
\ccsdesc[100]{Software and its engineering~Visual languages}

\keywords{\plainkeywords}

\maketitle

\section{Introduction}
Laser cutters are a key technology for personal fabrication~\cite{HCI-055} and likely the most popular machines available in Fab Labs. 
In the 2017 FabAcademy class, 80\% of the final projects made use of a laser cutter~\cite{fabacademy}.
Laser cutters work on vector data that defines where the material should be cut, and which parts should be engraved. 
Community platforms like \href{https://www.thingiverse.com}{thingiverse.com} or \href{https://www.grabcad.com}{grabcad.com} share numerous designs for a wide variety of projects.
These templates are commonly exchanged in the form of vector drawings and mostly use PDF, DXF, or SVG as a file format.

These templates, however, are usually made for a single type of material (e.g., $1/8$ inch plywood) and tweaked towards a specific type of laser-cutter. 
If an end-user downloads such a template to fabricate it from 3mm plywood, and a machine with a different kerf (i.e., the width of the cut), she has to adapt and tweak it to her specifications. 
Recent findings have shown editing digital models, such as SVG files, to be one of the biggest hurdles when using digital manufacturing technology~\cite{Mahapatra:2019jh}. 
Professional CAD environments, such as Rhino~\cite{rhino} or Autodesk Fusion 360~\cite{fusion360} support parametric design features to facilitate such changes to 2D or 3D designs. 
However, such features require designers (experienced users making the original model) and end-users (people using and adapting a designer's models) to use the same CAD environment. 
In practice, designers oftentimes use more advanced software environments and features that end-users do not want to be exposed to. 
Instead of sharing their parametric 3D model, designers therefore only share the 2D vector graphics drawings for laser cutting on platforms like Thingiverse. 
These vector graphics files do not require advanced tools but are static and lack the semantics of the underlying model, making it hard to adjust the design to new material or machine properties.

If an end-user, for example, downloads the SVG drawing of the miniature bench design shown in Figure~\ref{fig:walkthrough} and wants to manufacture it from a piece of material that is of different thickness than what the designer created it for, the end-user has to adjust eight rectangles (mortises) and 16 path segments (tenons).
Simple adjustments to laser cutter templates therefore often entail a large number of changes as all edges that correspond to material thickness need to be identified and adjusted. 
Similar adjustments are required to ensure a tight fit (i.e., press-fit) when fabricating a model with a laser cutter that has a different kerf~\cite{Roumen:2019}.



We propose \SysNoSpace, an extension to the SVG file-format that significantly facilitates creating and adapting parametric vector graphics designs for laser cutting. 
We will illustrate our contributions with an example.
A designer creates a vector drawing of the parts forming the miniature bench as shown in Figure~\ref{fig:walkthrough}a. 
In order to make it simpler for end-users to adjust the template to different material or machine parameters, the designer loads the SVG drawing into Inkscape\footnote{\url{http://www.inkscape.org}} if not already designing in this vector drawing application. 
The \Sys plug-ins offers various options to encode the additional information required to automate common changes when working with laser-cutter templates.
For example, the designer can specify which edges of the design correspond to the thickness of the material that is being used to build the bench. 
In our example, these are the height or width of the mortises in the side panels of the bench, and the length of the tenons. 
This information is stored in the SVG drawing.

To avoid end-users having to install additional software to adjust a template to their needs, \Sys templates embed a JavaScript library that turns the SVG file into a self-contained, responsive, and parametric template (Figure~\ref{fig:walkthrough}b). 
When an end-user opens a \Sys file in a web-browser (Figure~\ref{fig:walkthrough}c), this library renders a simple menu that allows her to alter common fabrication parameters to adjust the design to materials and machinery available to her.
The rendering of the model is updated accordingly and can easily be exported for fabrication either as a regular SVG file using the browser's ``save-as'' functionality or directly through the ``print'' menu.

\subsection{Contributions}
In this paper, we present three contributions:
\begin{enumerate}[noitemsep,topsep=0pt,parsep=0pt,partopsep=0pt]
    \item We introduce \SysNoSpace, a \emph{fully SVG-compliant file-format specification that encapsulates additional parametric features to facilitate automated adjustment for SVG models for laser cutting. } 
    \item We contribute a JavaScript library to turn the \Sys file-format into a \emph{self-contained, responsive, and parametric design template} that allows fine-tuning \Sys models in the web browser\footnote{Available as open source at \url{https://github.com/florianheller/lasersvg}}.
    \item We contribute a \emph{suite of Inkscape plug-ins to annotate traditional SVG files with parametric features and turn them into LaserSVG templates}\footnote{Available as open source at \url{https://github.com/florianheller/LaserSVG_Inkscape}}. 
\end{enumerate}

In the remainder of this paper we discuss the novel opportunities that come with our responsive laser-cutter file format, the workflow to make designs responsive, and an evaluation with real-world laser cutter templates.

\section{Related Work}
Laser cutters work on vector data that defines where the machine will cut or engrave things.
Designing objects to be manufactured with a laser-cutter, therefore, requires the creation of 2D vector drawings.
This process, however, is challenging \cite{Johnson:2012:SMS:2212776.2212390, Mahapatra:2019jh} and often leads to non-fitting results which then have to be re-adjusted, wasting time and material.  

Although being a 2D tool, the laser-cutter is also a very handy tool to create 3D structures which adds additional complexity to the design of the 2D shapes if these need to interlock properly. 
The research community developed two approaches to address this problem. 
The high-level approach takes 3D-models and breaks them down to components that can be manufactured on the laser-cutter, while the low-level approach directly works on 2D data.
Crdbrd \cite{CGF:CGF3037} and the work by Schwartzburg et al.~\cite{CGF:CGF12051}, for example, take a 3D model and convert it into planar sections.
Mesh2Fab~\cite{mesh2fab} assists in the disassembly of the 3D-models into components that can be manufactured from a target material, while Platener~\cite{Beyer:2015:PLF:2702123.2702225} replaces parts of the 3D-model with elements that can be laser-cut with the goal to speed up the production process compared to 3D-printing only.
These approaches can integrate material and machine parameters into their disassembly process.
However, changes in material thickness require the disassembly process to run again, while actually, only a series of edges need to be adapted in the 2D template.
CutCAD~\cite{Heller:2018:COT:3196709.3196800}, on the other hand, uses an inverse approach and starts from 2D shapes to create 3D objects by adding the required interconnections automatically. 

To simplify the common task of creating enclosures and boxes, we have scripts of variable complexity at hand. 
The \href{http://www.makercase.com}{MakerCase.com} web-tool creates simple rectangular boxes based on user-specified dimensions with parametric joints. 
The user can select whether the design should come with t-slot-joints, finger-joints, or without connections between the faces. 
The Boxes.py \cite{boxes.py} script provides more options regarding the shape of the resulting object and the joints.
For example, it is not limited to connections on the edge of a shape, but can also create mortise and tenon joints. 
Nevertheless, the user is limited to the functionality of the script, e.g., it does not support arbitrary shapes.
As such cases are often used as an enclosure for self-built devices, Weichel et al.~\cite{Weichel:2013:ECI:2460625.2460659} propose to take the bill of materials and the program code to generate a custom enclosure for a specific device.

Many projects also investigated the parameterization of designs, such that end-users can easily customize them to their needs.
A popular tool for designers is the Grasshopper plugin now included directly into the Rhino design environment, which allows designers to include algorithmic and parametric elements into their 3D models. 
Kyub~\cite{Baudisch:2019:KEM:3290605.3300796} is a 3D design environment tailor-made for laser-cutting which incorporates many best practices as tools to simplify the design. 
The Thingiverse Customizer~\cite{thingiverse} parses OpenSCAD scripts with specifically annotated variables and generates a graphical user interface for end-users to adjust these exposed variables. 
The adapted model can then be sent off for 3D-printing.
As writing OpenSCAD models is rather complex to understand for beginners~\cite{Yeh:2018:CMW:3173574.3174101}, other scripting approaches have been presented that build on the known knowledge of JavaScript~\cite{Kato:2017:FPD:3064663.3064681} and HTML~\cite{Yeh:2018:CMW:3173574.3174101}.
Both CraftML~\cite{Yeh:2018:CMW:3173574.3174101} and f3.js~\cite{Kato:2017:FPD:3064663.3064681} are designed to provide a low entry threshold to attract beginners to parametric design of 3D models, while at the same time providing a high ceiling so as to accompany designers up to professional level. 

Parametric approaches exist for two-dimensional models as well. 
Sketch-n-Sketch~\cite{Hempel:2016:SSP:2984511.2984575} is a script-based approach that stores the design as a lambda-calculus-program in the \emph{little} programming language, which the user can adjust using a direct manipulation interface to match its expected output.
Seamly2D~\cite{seamly2d} is a drawing application for responsive sewing patterns with the goal to make tailor-made clothing based on a series of body measurements.
Joinery \cite{Zheng:2017:JPJ:3059454.3059459} is a tool that designers can use to assign various types of joints to the edges in an SVG drawing. 
These projects, however, still require the use of specialized software to adjust the templates to end-users needs. 

Sharing designs is a core part of the Maker movement, but bare replication of the object can already be complicated due to a lack of information, context, or capabilities~\cite{Wakkary:2015:TAH:2702123.2702550}.
Furthermore, most of the shared laser-cutter templates consist of the flattened files, while sharing the underlying 3D model is uncommon for these projects~\cite{Roumen:2019}.
Therefore, it would be helpful to extend these currently static 2D exports with additional information, such that common parameters can easily be adapted.

SVG Parameters~\cite{svgSpecParam} and Parametric SVG~\cite{parametricsvg} introduce the use of variables and calculations in regular SVG files, making it possible to adjust their appearance.  
In \SysNoSpace, we extend this idea by making the templates responsive such that they can easily be adjusted without additional software, and by including functionality that specifically simplifies the creation and use of laser-cutter templates.

\section{Laser Cutter Templates}
To determine relevant features commonly used in laser cutter templates, we created a dataset of 40 real-world laser-cutter templates from Thingiverse. 
To generate a random set of models, we performed a search for the terms ``laser cut'' on Thingiverse in July 2020, which resulted in 7046 entries. 
From this corpus we created a sample by creating three sets of 30 models by taking every 235th entry from the search result, using a starting offset of 0, 50, and 100. 

%

From these 90 models, 13 only contained files for 3D printing, and 37 only used the laser-cutter in 2D, not for interlocking pieces, leaving 40 templates that build 3D-structures with interlocking pieces. 

We looked at the types of construction techniques that were used in these interlocking templates, with some templates combining multiple of these.
The three most common features were mortise and tenon-joints (23/40, 58\%), finger joints (13/40, 33\%), and slotting or lap joints (10/40, 25\%) as found in cross-section models. 
Two models used living hinges to bend material. 

We also looked at whether the template contains only paths to be cut, or if there are additional elements to be engraved. 
In total, 25 templates just contained cut information and 15 (37.5\%) used the cut and the engrave functionality of the laser-cutter.



\subsection{Material Thickness and Scale} 
\begin{figure}
	\centering
	\begin{overpic}[width=0.41\columnwidth]{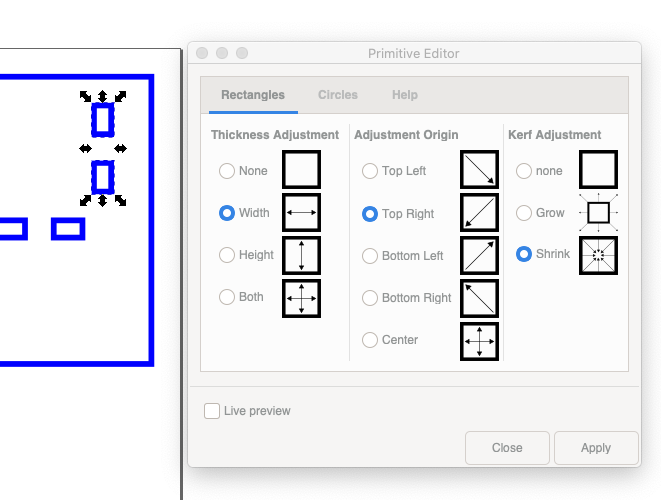} \put (5,5) {\textcolor{black}{\huge a)}} \end{overpic}
	\includegraphics[width=0.58\columnwidth]{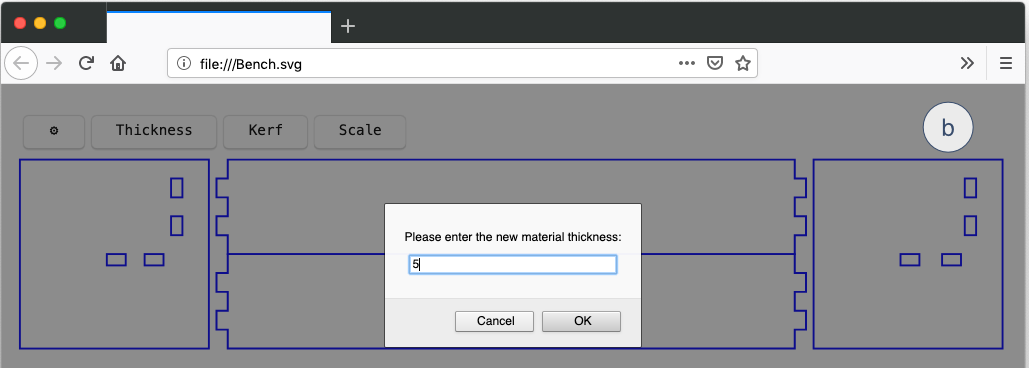}
	\caption{(a) In our editor, the designer can specify which dimension(s) of a rectangle correspond to material thickness, and whether the dimensions of the rectangle should compensate for the kerf-width. (b) When an end-user opens the template in the web-browser, she can adjust parameters such as material thickness, kerf-width, and scale to the properties of the materials and machinery at hand.}~\label{fig:basicFeatures}
    \vspace{-2.0em}
\end{figure}

A common adjustment for laser-cutter templates is to fit these to a new material thickness, for example, because the template is designed for a material thickness defined in inches and there is no exactly matching metric counterpart, or because the end-user just does not have the specified thickness at hand. 
Adjusting a laser-cut model easily leads to faulty designs and frustration: users often know what value they want to change (e.g., material thickness), but not how this change would reflect in the design.
To adapt a simple mortise and tenon joint as in the miniature bench example (Figure~\ref{fig:walkthrough}), the end-user has to consider which dimension of the mortise rectangle represents the material thickness. 
Finding all these thickness-dependent elements and adjusting them to other dimensions is a tedious and error-prone process as it requires building a geometrical understanding of the model. 
\Sys allows the designer of the template who created the object, or somebody else that has a clear concept of the object's geometry, to add this information to the file in order to simplify subsequent adjustments of the thickness parameter. 
As shown in Figure~\ref{fig:basicFeatures}a, the \Sys editor supports annotating thickness information in the SVG file-format.
The \Sys library embedded in the template uses this information to allow end-users to adjust the template to personal preference or available materials right in their web-browser (Figure~\ref{fig:basicFeatures}b).

These annotations also allow for a thickness-aware scaling function that leaves thickness-annotated elements untouched when scaling the SVG model to a different size while using the same material. 

\subsection{Kerf} 
Sharing laser-cutter templates also means that the item may be fabricated on a machine that is different from what the original design was made for. 
To make a template robust~\cite{doi:10.1002/qre.864, Roumen:2020}, i.e., universally adaptable to a wide variety of materials and laser-cutters, we need to consider the kerf: the width of the cut which depends on the material and on the type of laser-cutter machine~\cite{Roumen:2019,EffectOfProcessParametersonLaserCuttingProcess}. 
Compensating for kerf ensures the tightness of fit for so called ``press-fit'' joints that do not require the use of adhesives. 
Although the kerf of a laser cutter is thin, it still is about~0.2mm thick.
This means that when cutting a rectangular piece that is designed to be 100mm wide, the result will have a width of 99.8mm (0.1mm on every side). 

When using, for example, finnboard material, typically cut with a low resolution and high speed, the roughness of the cut leaves enough material for a solid connection making the kerf negligible. 
However, when cutting the same model from acrylic, which usually requires a higher resolution and slower speed, the edges will be much smoother resulting in a lack of friction to hold the pieces together. 

We introduce markers for designers to annotate the affected kerf edges in the \Sys editor (Figure~\ref{fig:basicFeatures}a). 
These annotations allow the \Sys model to shrink or grow these edges when end-users adjust the kerf parameter in the web-browser.

\subsection{Parametric Joints}
\begin{figure}
	\centering
	\begin{overpic}[width=0.405\linewidth]{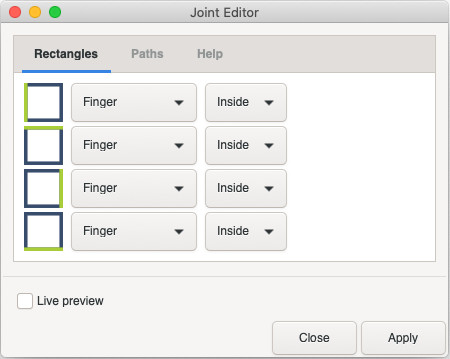} \put (5,5) {\textcolor{black}{\huge a}} \end{overpic}
	\begin{overpic}[width=0.59\columnwidth, trim=0 500 500 0, clip]{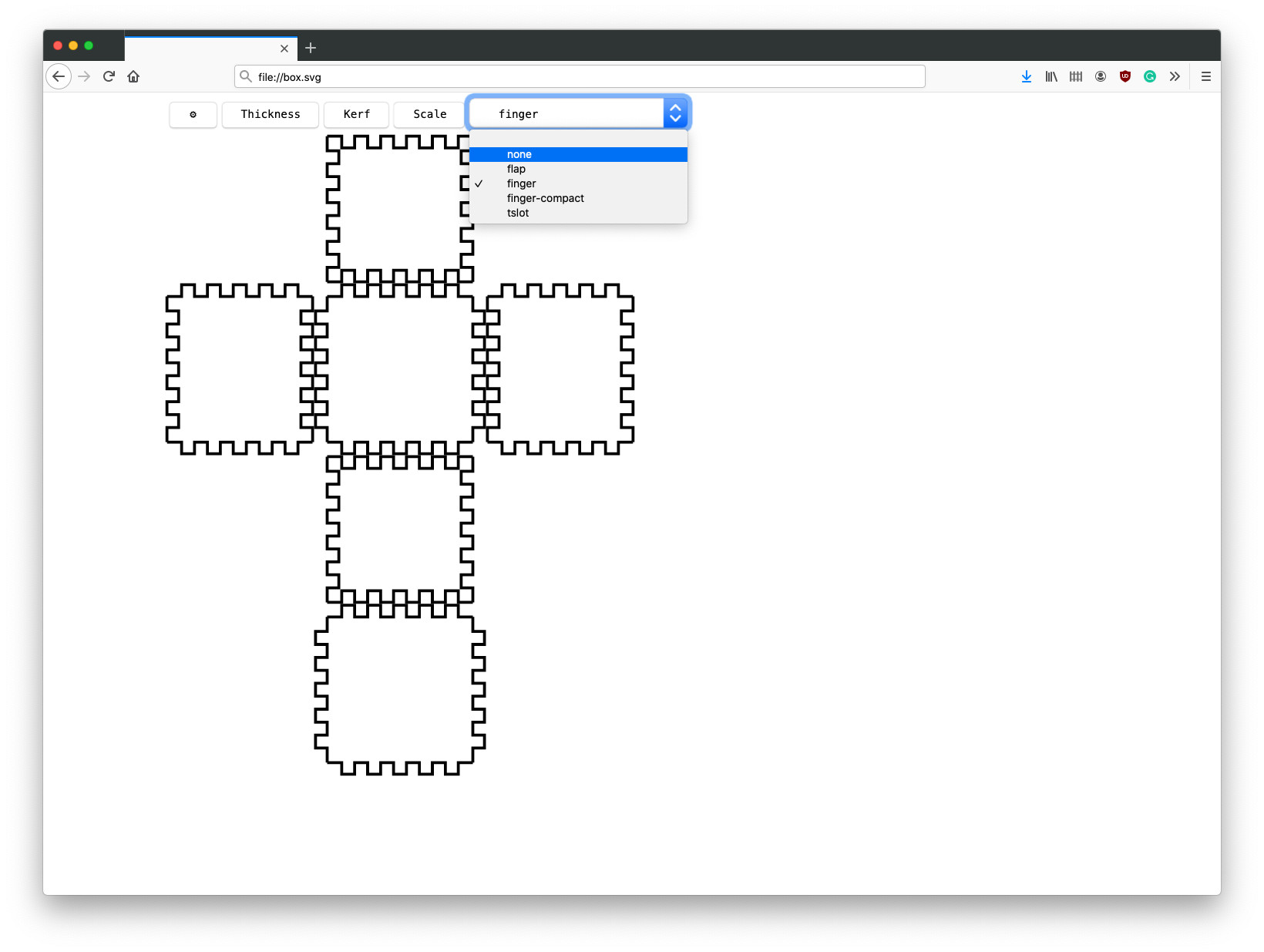} \put (10,5) {\textcolor{black}{\huge b}} \end{overpic}

	\caption{(a) The designer can assign a joint type and direction to the edges of a rectangle, which the \Sys library will render accordingly. (b) The end-user can change the type of joint through a simple drop-down menu to adjust the template to different materials or requirements.}~\label{fig:parametricjoints}
    \vspace{-2.0em}
\end{figure}

The design of an SVG model for laser cutting is a tedious process. For example, the creation of matching finger joints between all adjacent faces of a box (cf.~Figure~\ref{fig:examples}d). Furthermore, end-users might want to adjust the interconnection by changing the size or the number of fingers, or change the type of joint from finger-joint to a t-slot-joint to increase the stability of the object. \Sys therefore automates the creation of parametric joints of interchangeable type. Figure~\ref{fig:parametricjoints}a shows how a designer uses the \Sys editor to add a parametric joint to her design. End-users simply alter the type of joint using a simple menu available when opening the \Sys file in a web-browser (Figure~\ref{fig:parametricjoints}b).

\subsection{Laser Cutter Processing Options}
Laser cutter drivers use certain properties of the vector graphics elements to discern between those that need to be cut and those that are to be engraved.
The Trotec driver, for example, uses the outline color to determine which lines need to be cut, while the Epilog driver uses the stroke width for the same purpose. 
Therefore, \Sys introduces {\it cut} and {\it engrave} options for designers to label paths that require cutting or engraving respectively. 
In contrast to line color and thickness, these tags are not machine specific.
\Sys can apply manufacturer-dependent styling options to make a template compatible with the laser cutter at hand. 

Overall, similar control over the aforementioned parameters can be found when creating objects in parametric environments such as Rhino~\cite{rhino} or Kyub~\cite{Baudisch:2019:KEM:3290605.3300796}. 
However, these require sharing the 3D-model information, which is uncommon for laser cutting purposes~\cite{Roumen:2019}. Additionally, end-users have to use the same or an interoperable CAD environment as the one used by the designer.
In contrast, \Sys is fully compatible with the widely adopted SVG file-format, and allows end-users to adjust designs by simply opening them in a web-browser.

\section{The LaserSVG File Format}
\begin{figure}
	\centering
	\includegraphics[width=\columnwidth]{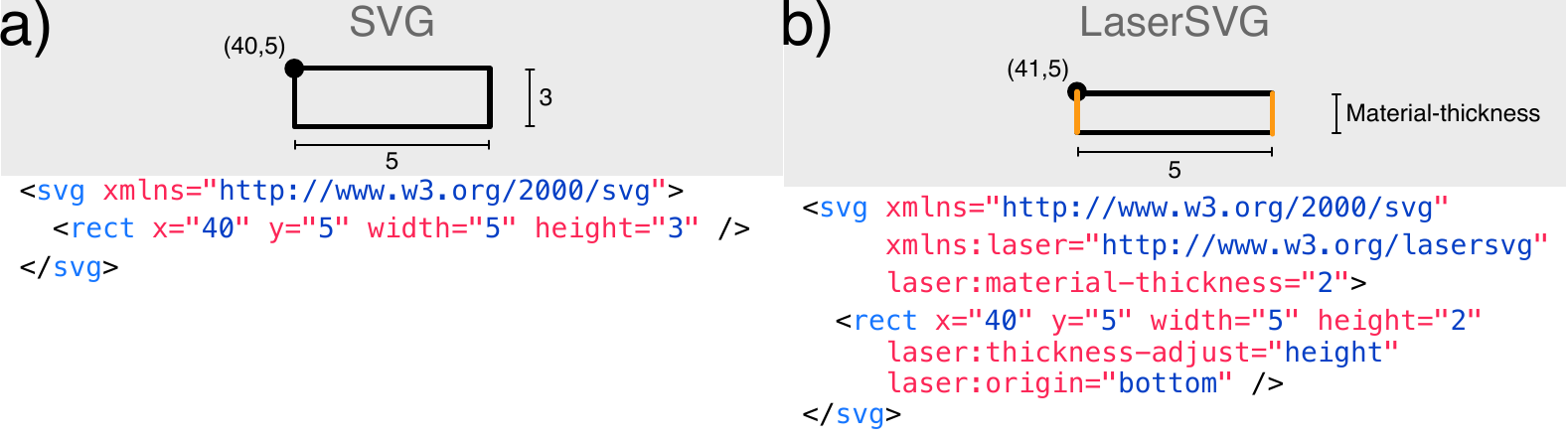}
	\caption{(a) Definition of a rectangle in SVG (b) The same rectangle resized to a new thickness based on the \emph{thickness-adjust} and \emph{origin} attributes defined in \Sys}~\label{fig:svgcode}
    \vspace{-2.0em}
\end{figure}

A main objective of \Sys is to integrate the features into an existing widely adopted file format, SVG, such that designers and makers can rely on the tools they already know and interoperability with fabrication tools is guaranteed.
Key to our approach is that the SVG-format ``allows inclusion of attributes from foreign namespaces on any SVG element''~\cite{svgSpec}. 
Additionally, SVG drawings can be scripted and made interactive using JavaScript~\cite{svgSpecProcessingModes}.
Even when an SVG viewer or editor does not support JavaScript or some custom SVG attributes, it is supposed to include these unknown attributes in the document object model (DOM) and preserve this information when exporting the SVG file.
This means that a \Sys file will be handled just like a regular SVG drawing by any software that does not implement our extension.
For example, if you load a \SysNoSpace-file into Inkscape, all attributes are kept and saved after the drawing has been changed.
Figure~\ref{fig:svgcode} shows respectively the SVG code for a basic rectangle (basic mortise joint) and \Sys code for a parametric version of that rectangle. The \Sys editor (see Section Editing Tools) facilitates the conversion from standard SVG to \SysNoSpace. 

All \Sys attributes and scripts can be included by adding \lstinline|xmlns:laser="http://www.w3.org/lasersvg"| 
to the opening \lstinline|<svg>| tag. All information that is included in addition to the standard SVG information is thus stored in the {\small {\ttfamily laser}} namespace. 
Our custom attributes are, therefore, prefixed with ``{\small{\ttfamily laser:}}''.

\subsection{Material Thickness}
To tag a specific edge as being of material thickness, we have to consider two cases: basic shapes and custom paths. 
For standard shapes like rectangles, circles, or ellipses, designers specify which dimensions correspond to material thickness by adding this dimension to the \emph{thickness-adjust} attribute of the shape (Figure~\ref{fig:svgcode}b). 
Similar to the \emph{resize} property in CSS3, this property takes {\it none}, {\it width}, {\it height}, and {\it both} as values.
In the SVG file format, rectangles are positioned by their top-left corner, which means that adapting the width or height results in changes towards the right and bottom respectively.
Therefore, \Sys makes it possible to add a resize \emph{origin} attribute to \emph{rect}-objects, which takes {\it right}, {\it bottom},  {\it bottom-right}, and {\it center} as values to scale the rectangle in the specified direction when changing parametric dimensions.
For more complex parametric shapes, designers can always switch to a more powerful custom path description.

\begin{figure}
	\centering
	\includegraphics[width=\columnwidth]{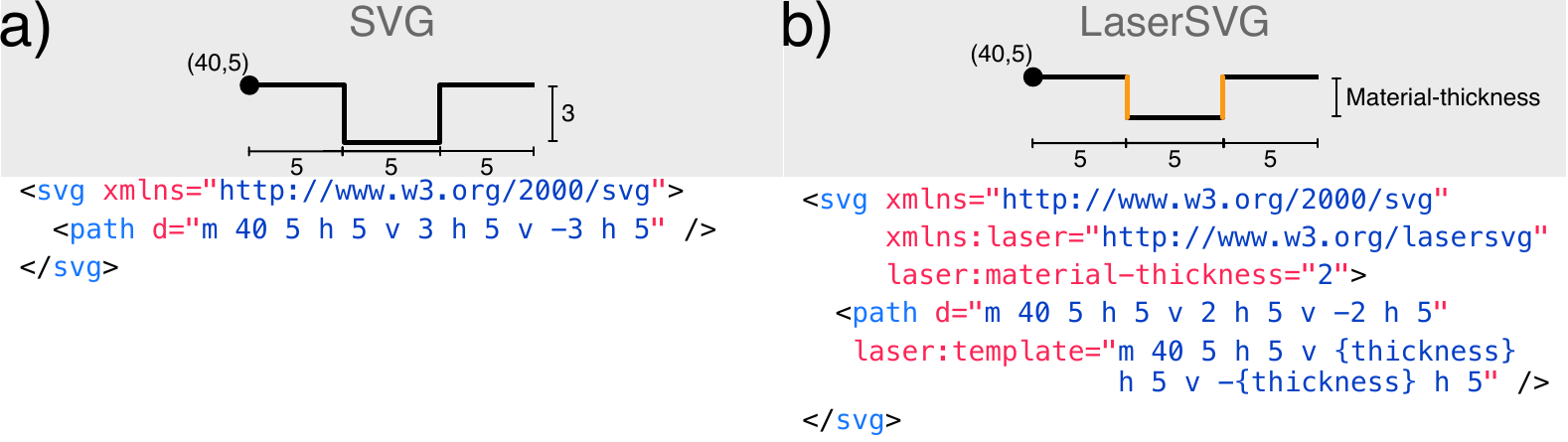}
	\caption{(a) Definition of a path in SVG (b) To resize the path, \Sys takes the \emph{template} attribute and generates a new path description by replacing the \{thickness\} labels with the new value.}~\label{fig:pathsvgCodePath}
    \vspace{-2.0em}
\end{figure}

Contour data for a path in SVG is stored in the \emph{``d''}-attribute of that path object.
This path description consists of a series of commands, including \emph{move (m)}, \emph{lineto (l)}, \emph{arc (a)}, and \emph{curve (c)}, which can either be specified in absolute (capital letter) or relative coordinates (lower letter).
As we want certain path segments to be parametric, \Sys adds a \emph{template} attribute which contains the same path description as the \emph{``d''}-attribute, but with the possibility to replace coordinates with \emph{\{thickness\}} labels for edges that are of material thickness (Figure~\ref{fig:pathsvgCodePath}).
Paths in \Sys therefore have two path descriptions, one with static data (``d'') and one augmented with parameters (``template''). This duplicate information ensures traditional SVG viewers and editors, not supporting \SysNoSpace, can still render and make changes to the file.
The \emph{\{thickness\}} label can also be extended with calculations within the curly brackets around the statement, such as \lstinline|{0.5 * thickness}| or \lstinline|{thickness + 2}|.
As the label can replace any coordinate in the path description, we can compensate for a thicker material by adjusting the origin of the path in the initial \emph{move} command (m) of the contour data.
Sometimes it is necessary to perform more complex calculations to keep an element at its place when changing the thickness, for example when working with triangular shapes whose center of rotation is not located at half of each dimension.
In between the curly braces the SVG file format also supports JavaScript code, such as variable assignments and access to mathematical functions. This allows for advanced parametric \Sys specifications, such as \lstinline|l{Math.cos(Math.PI/3)*thickness}| \lstinline|{Math.sin(Math.PI/3)*thickness}|. In contrast to assigning basic parameters to path segments and dimensions using point-and-click tools, advanced statements are entered by the designer through text fields available in the \Sys editor.
The thickness can be assigned globally through the \lstinline|material-thickness|-attribute in the root node of the SVG document, or, if needed, per path if the template is supposed to be fabricated using multiple materials. 

\subsection{Kerf Adjustment}
Similar to the material-thickness tags, \Sys introduces a kerf-adjust tag to allow for precise fits between laser cut joints.
For example, to ensure a press-fit between a mortise and tenon, the mortise should shrink and the tenon should grow by the kerf-width of the laser cutter. 
For basic shapes we introduce a \emph{kerf-adjust} SVG attribute which takes either {\it none}, {\it shrink}, or {\it grow} as value for the entire shape. 
Paths can have a {\small\path{kerf-mask}} SVG attribute, marking path segments that need to be adapted and the direction of that change.
The mask consist of one letter for each path segment, which can either be \emph{i} to ignore, \emph{g} to grow, or \emph{s} to shrink the corresponding path segment. 
Capital letters represent a full kerf-width while lower-case characters represent an adjustment by half the kerf-width.
For example, assigning the mask attribute \lstinline|"G i G i"| to a rectangular path with contour description \lstinline|"template =| \lstinline| M5 5 l0 10 l{thickness} 0 l0 -10 l-{thickness} 0"|, would grow the width by the kerf-width of the laser cutter. This results in a rectangle that is exactly 10mm high after laser cutting.
\begin{figure}
	\centering
	\includegraphics[width=\columnwidth]{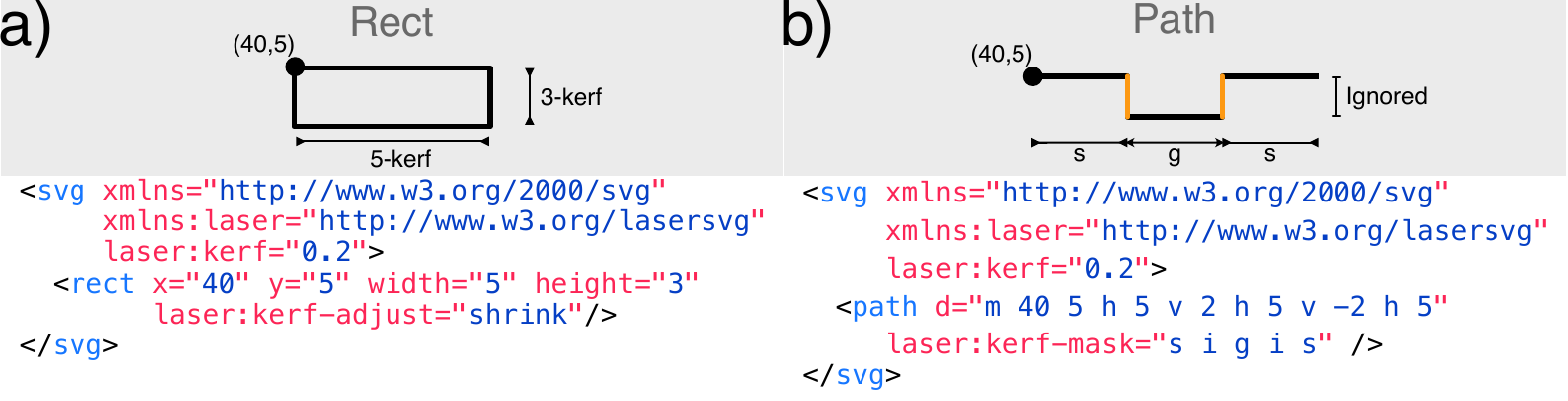}
	\caption{(a) Kerf adjustment attribute for a rectangle. (b) Kerf adjustment mask for paths. The initial move-command is ignored. Each letter represents the setting for the following path segment, where i=ignore, s=shrink, g=grow.}~\label{fig:pathsvgCodeKerf}
    \vspace{-2.0em}
\end{figure}





\subsection{Action}
\Sys introduces the \emph{action} attribute to label SVG elements for either cutting ({\it cut}) or engraving ({\it engrave}). A stylesheet ensures these SVG elements are then rendered according to the specifications of the laser cutter driver. 
Loading the stylesheet for a Trotec laser in the SVG file, for example, renders all elements labeled with {\it cut} in blue and elements to {\it engrave} will be filled black and outlined in red.
In contrast the Epilog stylesheet ensures {\it cut} elements are rendered with a thickness of .076~mm (hairline) while {\it engrave} elements are filled black.
Currently our stylesheet covers these two manufacturers, but it can easily be extended to support the specifics of other drivers. 
In case the properties conflict, end users can then switch between stylesheets by simply specifying the type of laser cutter at hand.


\subsection{Parametric Joints}
\begin{figure}[t]
	\centering
	\includegraphics[width=0.5\columnwidth]{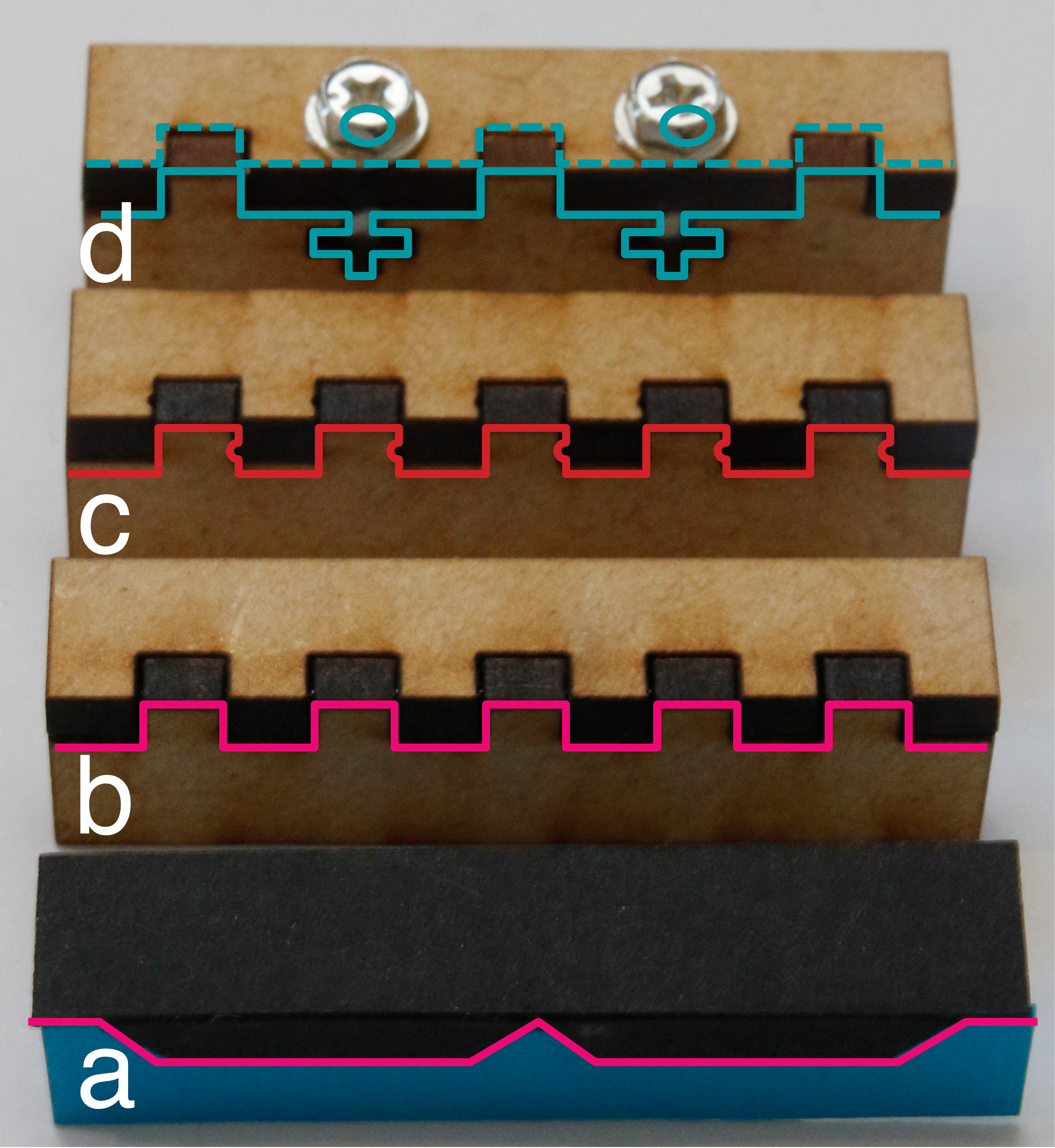}
	\caption{The different types of parametric joints we implemented in our \Sys library. (a) \emph{flap}-joint (b) \emph{finger}-joint (c) \emph{compact finger}-joint for additional grip (d) \emph{t-slot}-joints (inside \& outside)}~\label{fig:parametricJoints}
    \vspace{-2.5em}
\end{figure}

\Sys introduces higher-level attributes to SVG path elements that replace an edge with a specific joint. 
This allows end-users to change the type of joint, while avoiding the tedious design skills that are required to do this manually. To support parametric joints, several attributes are available in \Sys:
The \emph{joint} attribute assigns an ID to the edges of two elements that join together. 
This allows the \Sys editor to keep parameters in sync and ensures, for example, that the two edges have the same joint-type.
The \emph{joint-type} attribute specifies the type of joint that should be rendered instead of the original edge. Supported values in the current \Sys version include (compact) finger-joints, t-slot-joints, and flaps (cf.~Figure~\ref{fig:parametricJoints}). 
The \emph{joint-direction} attribute specifies which side of a joint is drawn.
For example, in the case of a finger-joint, the \emph{outside} direction adds the fingers to the edge of a face while the \emph{inside} direction provides the corresponding notches. 
As rectangular shapes are frequently used in SVG models, \Sys offers the possibility to define a joint type for each side, e.g., \emph{joint-top-type} and \emph{joint-top-direction}, which avoids requiring the designer to convert the rectangle into single paths.
All generated joints paths also come with a \Sys \emph{template} description, embedding thickness tags for automated adjustments as explained above.

Table \ref{tab:commands} offers a detailed overview of all \Sys attributes and whether they can be assigned globally in the root node or per SVG element. 
Some of the attributes can easily be changed by the end-user as we will describe in the following section.

\begin{table}
\centering
  \caption{LaserSVG attributes and their potential scope, i.e., whether it applies to the entire template or just a single element. Gray cells are parameters an end-user can quickly adjust through the menu in the web-browser.}
  \label{tab:commands}
  \begin{tabular}{r|c|c}
    \thead{Attribute} & \multicolumn{2}{c}{\thead{Scope}} \\
    & Element & Global \\
    \midrule
    {\ttfamily material-thickness} & \checkmark & \cellcolor{light-gray}\checkmark\\
    {\ttfamily kerf} &  &  \cellcolor{light-gray}\checkmark\\
    {\ttfamily scale} &   & \cellcolor{light-gray}\checkmark \\\hline
    {\ttfamily origin} & \checkmark   \\
    {\ttfamily kerf-adjust} & \checkmark   \\\hline
    {\ttfamily joint-type} & \checkmark & \cellcolor{light-gray}\checkmark \\
    {\ttfamily joint-direction} & \checkmark  \\
    {\ttfamily joint} ID & \checkmark  \\\hline
 	{\ttfamily action} &  \checkmark & \checkmark\\
    {\ttfamily action-value} & \checkmark & \checkmark\\
\end{tabular}
\end{table}

\section{Self-Contained Responsive \Sys templates}
The \Sys file-format is an extension to the standard SVG format and does not require end-users to install additional software to make adjustments to SVG models for laser cutting.
Instead, a \Sys template is a self-contained, responsive format that, when opened in a standard web-browser, renders GUI elements to adjust the contained SVG model.
Therefore, we implemented a JavaScript library that is included in any \Sys file and implements two dynamic functions. 
First, it renders a dynamic menu that appears if the SVG file is opened directly in a SVG renderer that supports JavaScript, i.e., a regular web-browser (see Figure~\ref{fig:basicFeatures}b). 
These menu items offer features to adjust parameters in the current \Sys file. 
This end-user menu is not shown when the designer opens the \Sys file in the \Sys editor or a traditional SVG editor. 
Second, the library interprets all ``{\small{\ttfamily laser:}}'' attributes and ensures the SVG model is rendered and updated according to the set parameters, such as adaptive thickness, thickness-aware scaling, kerf-adjustment, and parametric joints.

\section{\Sys Editor}
\begin{figure}
	\centering
	\begin{overpic}[width=0.3\linewidth]{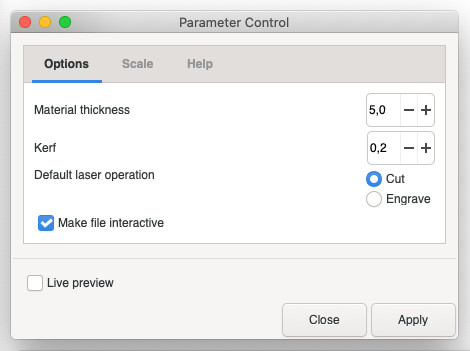} \put (5,5) {\textcolor{black}{\huge a}} \end{overpic}
	\begin{overpic}[width=0.3\linewidth]{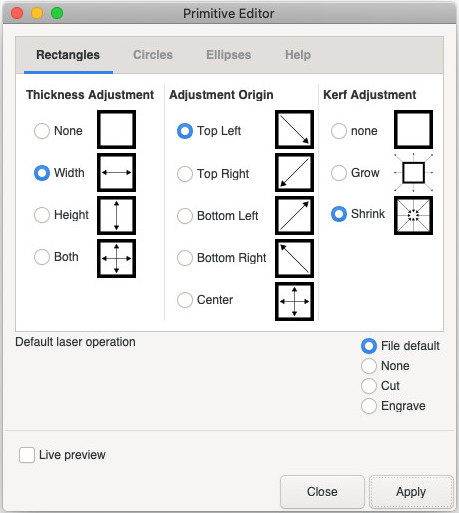} \put (5,5) {\textcolor{black}{\huge b}} \end{overpic}
	\begin{overpic}[width=0.3\linewidth]{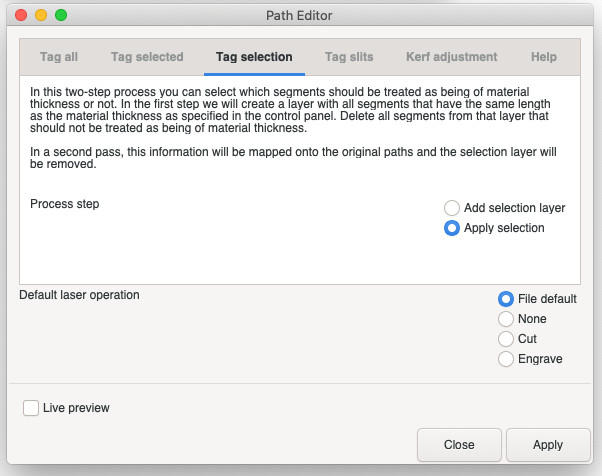} \put (5,5) {\textcolor{black}{\huge c}} \end{overpic}
	\caption{The \Sys Inkscape plug-ins: (a) In the control window, the user can set the material thickness and kerf values, as well as a default operation for elements of the drawing. If needed, the drawing can also be scaled by respecting the thickness-labels. 
	(b) the editor for geometric primitives provides options for the dimensions in which to adapt to material thickness, as well as the origin relative to which this change is applied. If needed, a kerf-compensation can be set. 
	(c) The path editing tool provides a number of helper functions to handle the various parameterizations of paths.}~\label{fig:Editor_Path}
\end{figure}

In this section we will describe how our suite of Inkscape plug-ins supports the designer in annotating the drawing with the additional \Sys tags to turn it into a parametric template. 
As Inkscape's plug-in API only provides a limited set of user interface elements and no interactivity in the plug-in windows, we decided to split functionality over several panels.
To turn the current drawing into a \Sys template, the designer starts by specifying the material thickness the model is designed for, a default kerf-width, and whether the template should be responsive, i.e., by embedding our JavaScript library (Figure~\ref{fig:Editor_Path}a). 
Next, the designer selects individual elements, such as paths, segments, and edges, and assigns parameters using simple menu options.
For basic shapes, this can be done through the configuration panel for geometric primitives (Figure~\ref{fig:Editor_Path}b).
For custom paths, we provide a series of helper functions to simplify the conversion of the text-based path-description into a path template (Figure~\ref{fig:Editor_Path}c).

The basic helpers work as follows: 
\emph{Tag all} parses all segments of the selected paths and tags those that are the same length as the specified material thickness. 
\emph{Tag selected} uses the selection from Inkscape's node editing tool, and tags the segments between adjacent selected nodes. 
This helper is useful if only a few segments have to be tagged and allows the designer to work with the tools she is already accustomed to. 
\emph{Tag selection} is a two-step process. In the first step, the tool checks the length of each segment in the selected paths. 
If the length of a segment matches the specified material thickness, our tool creates a highlight of that segment in a separate layer. 
The designer then checks these highlights and deletes those that are not supposed to be tagged. 
Once the selection is complete, the tool tags the specified selection in the original path description to create the template. 

All these helper functions take the angle of the path segments into consideration and add the necessary calculations into the tag.

\subsection{Additional Support}
\begin{figure}[htp]
	\centering
	\includegraphics[width=0.3\linewidth]{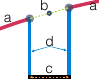} 
	\caption{To keep a slit centered while adjusting the material thickness, the slotting tool in the \Sys editor parameterizes the drawing by extending or shortening the edges leading from and to the slit (red) along the line connecting their start and end point respectively. The bottom segment (black) represents material thickness and the parallel blue lines are shortened or extended accordingly.}~\label{fig:slotting}
\end{figure}

Our initial evaluation of laser-cutter templates showed that slotting joints, as used in cross-section models, are a common feature.
Simply specifying the width of the slot as being of material thickness may have unwanted repercussions in the model as the overall length changes with thickness. 
In most cases, adapting the width of a slot should keep its position stable and therefore affects not only the width of the slot bottom, but also the length of the two path segments leading to and away from it (Figure~\ref{fig:slotting}a).
If we increase the size of the slot in the example from Figure~\ref{fig:slotting}, the length of the blue segments needs to be adjusted as well. 
\Sys supports this through its calculation feature, but requires to transform several path commands into parametric variants. 

Similar to the \emph{Tag selection} option, the user can select the \emph{Tag slits} option in the path editing panel (Figure~\ref{fig:path_Panel}) 
This is again, a two-step function, which in the first step highlights all segments in the selected path that are the size of the specified material thickness. 
The designer can then delete the highlights that were erroneously considered to be slit bases. 

First, we calculate the center point of the line connecting the two edge points of the slot (Figure~\ref{fig:slotting}b). 
The enclosing segments will be extended or shortened along this line. 
We then adjust the length of the bottom segment (Figure~\ref{fig:slotting}c) to fit the material thickness, and finally, adjust the length of the two parallel segments forming the slot (Figure~\ref{fig:slotting}d) if needed, i.e., the adjacent segments and the slit base are not parallel.


\section{Evaluation}
To evaluate our system, we tried to convert the designs of the 40 existing models from our random sample into LaserSVG templates. 
This \textit{post-hoc} conversion is somewhat unusual as it requires a third person to dive into the geometric reasoning of the model. 
In a real-world scenario, the designer would do that while designing, or in a post-processing step of the template, making the task simpler.

The effort necessary to convert an existing design highly depends on the quality of the file. 
The templates in our dataset come in a variety of file formats generated from various drawing programs.
Even if the template author provides an SVG drawing, its quality highly depends on the origin of the file. 
Some files consist of single lines only, meaning they do not contain any coherent shape information. 
Others do not make use of primitives, but express every shape as path, although using the primitives would be simpler. 
The paths often contain a number of unnecessary nodes, leading to zero-length segments in the path, or the paths approximate curves using a large number of straight lines. 
Furthermore, Inkscape loads files with pixels as default unit, meaning that the values in the path descriptions do not correspond to the actual values in mm.
If the author of the template did not specify the material thickness the design was made for, it is hard to determine wether the template already accounts for kerf. 
Some templates are not even consistent within their drawing, making the editing process unnecessary complicated (e.g., Thingiverse ID \thing{757581}). 
Therefore, most templates require some pre-processing to make the templates usable.
This pre-processing includes decluttering of the file, conversion to mm, scaling to specified dimensions, and decluttering or joining of paths into meaningful units. 
This step is not directly related to the conversion into a \Sys template, but would be required in similar form for any modification of the given design and sometimes even for the mere use of the template.
Depending on the export quality, it is sometimes simpler to redraw part of the design as, for example, dwg-files often only contain single lines which all have to be adjusted in case one wants to change something.

In the following we will discuss our observations 

For very simple models with only mortise \& tenon-joints, it is sufficient to specify the direction of the thickness adaptation and the anchor point for that change (cf.\ Thingiverse IDs \thing{2913091}, \thing{4056240}).

In many cases, the tenon is designed to match the material thickness to form a nicely closed surface (Thingiverse IDs \thing{149}, \thing{23953}, \thing{124818}, \thing{1906241},\thing{2442411}, \thing{2839796}, \thing{2975003}, \thing{3016694}, \thing{3897974}). 
which can easily be achieved by using the ``\textit{Tag selection}'' functionality from our plug-in. 

In enclosures for electronics, such as the popular cases for the RaspberryPi (\thing{27597},\thing{565828},\thing{566678}), keeping the inside dimensions is crucial for the connectors to stay centered in their designated cutouts. 
For the outsides of the material to stay aligned, some edges have to be elongated. 
When editing the file manually, one would simply add a calculation such as \lstinline|{24+thickness}|. 
For graphical editing, we present two solutions: 
One can simply add an additional node into the line which is exactly \textit{thickness} away from the next one.
This segment can then be tagged to be adjusted along with the remaining segments using the ``\textit{Tag selection}'' tool. 
The second option is to compose the calculation through a graphical user-interface in the ``\textit{Tag selected}''-panel, by specifying the amount that has to be added. 

The other common interconnection between plates are finger joints. 
If the inner dimensions of the enclosure are to be kept constant, the fingers can easily be tagged using the ``\textit{Tag selection}'' tool (\thing{946265}, \thing{1169378}, \thing{2005219}, \thing{2281214}, \thing{3016694}, \thing{3140483}, \thing{3320769}, \thing{3419311}). 
If the outer dimensions are to stay constant (\thing{9111}, \thing{2994225}), we need to compensate for that by deepening the cuts between the fingers. 
For the edge formed by the ends of the fingers to remain at the same position, we need to subtract \textit{thickness} from the edge leading to the finger joint. 
This can, again, be done manually or through the ``\textit{Tag selected}''-panel of the path editing plug-in. 

Finally, slot-joints (\thing{4573}, \thing{24888}, \thing{25864}, \thing{59216}, \thing{1340543}, \thing{3015485}) are a common way to interconnect pieces in planar section models (\thing{757581}, \thing{3127666}, \thing{3906772}, or to create non-cubic enclosures (\thing{2017833}). 

These are the most complicated to parameterize, as not only the slot-base needs to be adjusted, but the edges left and right of the slot as well. 
We have added tool-support for this kind of connection (see \autoref{fig:slotting}).
Unfortunately, while this tools works in a number of cases (\thing{2017833}, \thing{3015485}), it does not work satisfactorily in  complex examples such as the rabbit (\thing{757581}) or the basket (\thing{3127666}).

\section{Limitations}

During the evaluation we encountered some situations which required workarounds in Inkscape or, in rare cases, manual calculations. 
In all cases, this was not a limitation by the \Sys library, but by the editing tools which we will improve based on these findings.

One interesting case was Thing no. \thing{2442411}, which looks simple, but it contains a stack of three layers of material with the two outer layers being held by mortises. 
This means that these two mortises not only have to be of material thickness, but also need to be spaced by the same amount. 
While it is easily possible to achieve this by adding a calculation to the initial \textit{move}-command of the path, there is currently no tool support to specify this behavior graphically.

If mortises are not rectangular, but rounded rectangles defined as paths such as in thing \thing{2281214}, it is currently not possible to tag these automatically. 
As this is a limitation of the editing tool, it is possible to manually adjust the size of the straight segment between the rounded corners. 
The better solution is to define these as rectangles instead of paths, which then provides full support.

A possible workaround for the slot-joints in complex models is to use an approach demonstrated in thing \thing{1340543}, where instead of defining a closed path including the slots, one just draws the outline of the piece and overlays these with rectangles at the location of the slots. 
While the files are optically different, after cutting the design the result is identical.

The rocket \Sys template embeds more advanced mathematical statements to properly use the material-thickness parameter in the tri-wing cutouts. These statements ensure the tri-wing cutouts are always centered in the circular pieces while changing parameters or scaling the model.
We converted a simple rocket \cite{rocket} template using relative path descriptions and the calculation features. 
The challenge are the non-orthogonal edges in the circular elements that need to be adjusted to the material thickness.
We added $sin$ and $cos$ factors to the calculation, and adjusted the starting point of the tri-wing cutout by using a thickness-dependent offset to keep it centered within the circles.
We implemented and embedded a JavaScript function to perform this calculation based on length and thickness directly in the template. 
The \emph{template} attribute calls this function to adjust the initial \emph{move}-command in the path description \lstinline|M{24-0.5*thickness}{32-triangleRadius(9,thickness)}|.
Calculating the center of rotation for the triangular shape makes this example quite complex. However, it shows the high ceiling that scripting in \Sys offers.

\begin{figure}
	\centering
	\includegraphics[width=\columnwidth]{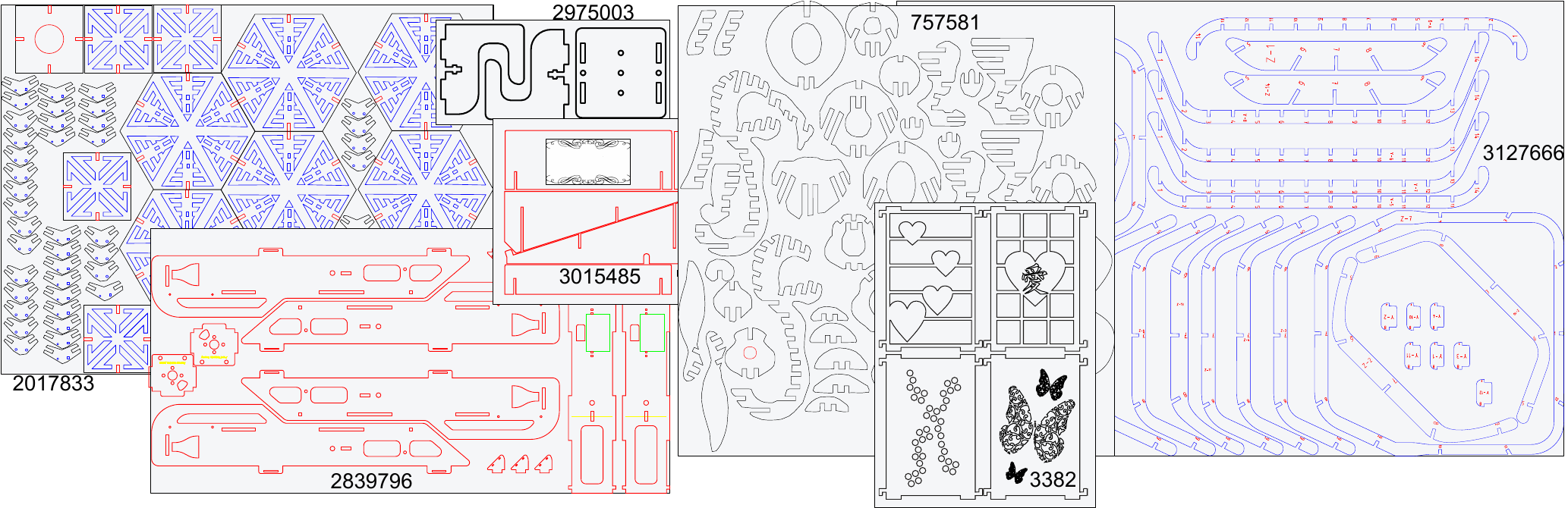}
	\includegraphics[width=0.3\columnwidth]{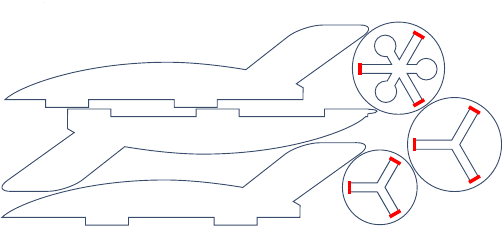}
	\caption{\textbf{ top:} examples from our evaluation dataset with their respective Thingiverse ID. Our tools were not sufficient to successfully parameterize the complex cross-section models \thing{757581} and \thing{3127666}.
	\textbf{ bottom:} To keep the tri-wing cutout centered, we used the scripting functionality to calculate its circumference and center-point.}~\label{fig:examples}
    \vspace{-2.0em}
\end{figure}


\section{Conclusion \& Future Work}



In this paper, we presented \Sys, a fully SVG compliant file format to ease the process of adjusting vector models for laser cutting. 
\Sys files are self-contained and responsive and therefore allow end-users to make adjustments to the SVG model by simply opening the file in the web browser. 
As such, end-users are not required to install and use CAD modeling software: a web-browser suffices for updating the vector model, either interactively or automatically. 
We also contribute a \Sys editor to facilitate the process for designers to convert regular SVG files to \Sys templates. 
While \Sys is very powerful, our approach also has limitations which reveal several interesting challenges for future research.


First, ensuring vector drawings are always coherent when adjusting parameters can be complicated for advanced designs. 
For example, ensuring that intricate designs, such as the rocket (Figure~\ref{fig:examples}) scales appropriately requires advanced mathematical statements. 
Similarly, supporting complex parametric joints along Bezier curves requires formulating the positions of all control points in relation to the material thickness. 
More research is needed on how to support designers in specifying complex mathematical statements and dependencies in vector drawings.


Second, although \Sys files are backwards compatible and can be rendered and adjusted with traditional SVG editors, making significant adjustments, such as adding control points to a path requires annotating the vector design again for conversion to \SysNoSpace. 
Designers also tend to make SVG designs starting from 3D models. 
Making changes to the 3D model and exporting them again will overwrite all \Sys annotations. 
In the future, we envision the adoption of the \Sys format in more design environments. 
This would also allow directly linking parameters of 3D solid models to \Sys parameters and integrate \Sys in the designers workflow without requiring additional conversions or annotations.
Using a direct export also reduces the complexity of the SVG files, which, depending on how many conversion steps there have been, can become quite ill-formed.

Last, \Sys can be extended with additional parametric features, such as automatic adjustment of living hinges when using a different material thickness and interactive selection of different joints that are a better fit for changing kerf sizes.

\bibliographystyle{ACM-Reference-Format}
\bibliography{laserSVG}

\end{document}